\documentclass[twocolumn]{aastex631}

\usepackage{ulem}
\usepackage{amsmath}
\usepackage{booktabs}
\usepackage{epstopdf}
\def\meth {CH$_3$OH}
\def\hho  {H$_2$O}

\defcitealias{2019ApJ...885..131R}{R19}

\def\kms  {km~s$^{-1}$}
\def\masy {mas~y$^{-1}$}

\def\deg  {\ifmmode {^\circ}\else {$^\circ$}\fi}

\def\p    {\phantom{0}}
\def\d    {\ifmmode {{\rlap{.}}^\circ}\else {${\rlap{.}}^\circ$}\fi}
\def\s    {\ifmmode {{\rlap{.}}^s}\else {${\rlap{.}}^s$}\fi}
\def\as   {\ifmmode {{\rlap{.}}^{''}}\else {${\rlap{.}}^{''}$}\fi}
\def\HI   {H~{\small I}}
\def\HII  {H~{\small II}}
\def\chisqpdf {\ifmmode {\chi^2_{\rm pdf}}\else {$\chi^2_{\rm pdf}$}\fi}
\def\chisq    {\ifmmode {\chi^2}\else {$\chi^2$}\fi}
\def\pa    {\ifmmode {\psi} \else {$\psi$}\fi}

\newbox\grsign \setbox\grsign=\hbox{$>$} \newdimen\grdimen \grdimen=\ht\grsign
\newbox\laxbox \newbox\gaxbox
\setbox\gaxbox=\hbox{\raise.5ex\hbox{$>$}\llap
{\lower.5ex\hbox{$\sim$}}}\ht1=\grdimen\dp1=0pt
\setbox\laxbox=\hbox{\raise.5ex\hbox{$<$}\llap
{\lower.5ex\hbox{$\sim$}}}\ht2=\grdimen\dp2=0pt

\def\vlsr  {\ifmmode {v_{\rm LSR}}\else {$v_{\rm LSR}$}\fi}
\def\Vlsr {\ifmmode {V_{\rm LSR}} \else {$V_{\rm LSR}$} \fi}
\def\Vlsra {\ifmmode {{V_{\rm LSR}}^{a}} \else {$V_{\rm LSR}^{\ast}$} \fi}
\def\vhelio{\ifmmode {v_{Helio}}\else {$v_{Helio}$}\fi}

\def\Ge {G019.60$-$00.23}
\def\Gf {G020.08$-$00.13}
\def\Ga {G033.64$-$00.22}
\def\Gb {G035.57$-$00.03}
\def\Gc {G041.15$-$00.20}
\def\Gd {G043.89$-$00.78}

\def\To    {\ifmmode{\Theta_0}\else{$\Theta_0$}\fi}
\def\Ro    {\ifmmode{R_0}\else{$R_0$}\fi}
\def\Ts    {\ifmmode{\Theta_s}\else{$\Theta_s$}\fi}
\def\Tdot  {\ifmmode{d\Theta\over dR}\else{$d\Theta\over dR$}\fi}

\def\ura   {\ifmmode {\mu_\alpha}\else {$\mu_\alpha$}\fi}
\def\udec  {\ifmmode {\mu_\delta}\else {$\mu_\delta$}\fi}

\def\ul    {\ifmmode {\mu_l}\else {$\mu_l$}\fi}
\def\ub    {\ifmmode {\mu_b}\else {$\mu_b$}\fi}

\def\uml   {\ifmmode {v_{gr}}\else {$v_{gr}$}\fi}
\def\umb   {\ifmmode {v_b}\else {$v_b$}\fi}
\def\vsrad {\ifmmode {v_{rad}}\else {$v_{rad}$}\fi}

\def\upl   {\ifmmode {v^p_{gr}}\else {$v^p_{gr}$}\fi}
\def\upb   {\ifmmode {v^p_b}\else {$v^p_b$}\fi}
\def\vprad {\ifmmode {v^p_{rad}}\else {$v^p_{rad}$}\fi}

\def\Vo    {\ifmmode {V^{Std}_\odot}\else {$V^{Std}_\odot$}\fi}
\def\Uo    {\ifmmode {U^{Std}_\odot}\else {$U^{Std}_\odot$}\fi}
\def\Wo    {\ifmmode {W^{Std}_\odot}\else {$W^{Std}_\odot$}\fi}

\def\V     {\ifmmode {V_\odot}\else {$V_\odot$}\fi}
\def\U     {\ifmmode {U_\odot}\else {$U_\odot$}\fi}
\def\W     {\ifmmode {W_\odot}\else {$W_\odot$}\fi}

\def\Vs    {\ifmmode {V_s}\else {$V_s$}\fi}
\def\Us    {\ifmmode {U_s}\else {$U_s$}\fi}
\def\Ws    {\ifmmode {W_s}\else {$W_s$}\fi}

\shortauthors{Bian et al.}

\begin{document}

\title{On the Structure of the Sagittarius Spiral Arm in the Inner Milky Way}
\author{S. B. Bian}
\affiliation{Purple Mountain Observatory, Chinese Academy of Sciences,
Nanjing 210033, China; xuye@pmo.ac.cn}
\affiliation{School of Astronomy and Space Science, University of
Science and Technology of China, Hefei 230026, China}
\author{Y. W. Wu}
\affiliation{National Time Service Center,
Key Laboratory of Precise Positioning and Timing Technology,
Chinese Academy of Sciences, Xi'an 710600, China}
\author{Y. Xu}
\affiliation{Purple Mountain Observatory, Chinese Academy of Sciences,
Nanjing 210033, China; xuye@pmo.ac.cn}
\affiliation{School of Astronomy and Space Science, University of
Science and Technology of China, Hefei 230026, China}
\author{M. J. Reid}
\affiliation{Center for Astrophysics~$\vert$~Harvard $\&$ Smithsonian,
60 Garden Street, Cambridge, MA 02138, USA}
\author{J. J. Li}
\affiliation{Purple Mountain Observatory, Chinese Academy of Sciences,
Nanjing 210033, China; xuye@pmo.ac.cn}
\author{B. Zhang}
\affiliation{Shanghai Astronomical Observatory, Chinese Academy of
Sciences, Shanghai 200030, China}
\author{K. M. Menten}
\affiliation{Max-Planck-Institut f$\ddot{u}$r Radioastronomie, Auf
dem H{\" u}gel 69, 53121 Bonn, Germany}
\author{L. Moscadelli}
\affiliation{INAF-Osservatorio Astrofisico di Arcetri, Largo E. Fermi
5, 50125 Firenze, Italy}
\author{A. Brunthaler}
\affiliation{Max-Planck-Institut f$\ddot{u}$r Radioastronomie, Auf
dem H{\" u}gel 69, 53121 Bonn, Germany}

\begin{abstract}

We report measurements of trigonometric parallax and proper motion for
two 6.7 GHz methanol and two 22 GHz water masers located in the far portion of the Sagittarius spiral arm as part of the BeSSeL Survey. 
{Distances for these sources
are estimated from parallax measurements combined with 3-dimensional
kinematic distances. 
The distances of G033.64$-$00.22, G035.57$-$00.03, G041.15$-$00.20, and G043.89$-$00.78 are $9.9\pm0.5$, $10.2\pm0.6$, $7.6\pm0.5$, and $7.5\pm0.3$ kpc, respectively.}  Based on these 
measurements, we suggest that the Sagittarius arm segment beyond about 8 kpc from the Sun in the first Galactic quadrant should be adjusted radially outward relative to previous models. This { supports the suggestion of Xu et al. (2023) that} the Sagittarius and Perseus spiral arms { might} merge in the first quadrant before spiraling inward to the far end of the Galactic bar.

\end{abstract}


\section{Introduction} 
\label{sect-intro}

Over the last 16 years, the parallaxes and proper motions of over 200 masers associated with high-mass star-forming regions (HMSFRs) have been measured (\citealt{2019ApJ...885..131R}, hereafter \citetalias{2019ApJ...885..131R}; \citealt{2020PASJ...72...50V}), which trace the spiral arms and three-dimensional (3D) motions of their young stars.
While the nearby spiral arms of the Milky Way have been mapped in
detail \citep[e.g.,][]{2016SciA....2E0878X}, there are few parallax 
measurements for sources with distances beyond $\approx$10 kpc. In order to extend our mapping of the Milky Way, in this work we report parallax measurements of four distant maser sources{, G033.64$-$00.22, G035.57$-$00.03, G041.15$-$00.20, and G043.89$-$00.78,} located past the tangent point of the Sagittarius spiral arm in the first Galactic quadrant. 
{ In addition, in order to better constrain the spiral arm structure in distant regions, we also calculate the 3D kinematic distances for several sources whose distances are greater than about 8 kpc. \citet{2022AJ....164..133R} has shown 3D kinematic distances to be generally superior to parallax distances for sources well past the Galactic center.  }

\section{Observations and Data Analysis} \label{sect-obser}

\begin{deluxetable*}{lllrrrcl}[t]
	\tabletypesize{\footnotesize}
	\tablecaption{Positions and Brightnesses \label{table:positions}}
	\tablehead{
		\colhead{Source} & \colhead{R.A. (J2000)} &
		\colhead{Dec. (J2000)} & \colhead{$\Delta\Theta_{\rm E}$} & \colhead{$\Delta\Theta_{\rm N}$} & \colhead{Brightness}
		& \colhead{$V_{\rm LSR}$} & \colhead{Beam} \\
		\colhead{} & \colhead{$\mathrm{(^h\;\;\;^m\;\;\;^s)}$}
		& \colhead{$(\degr\;\;\;\arcmin\;\;\;\arcsec)$} & \colhead{($^{\circ}$)} & \colhead{($^{\circ}$)}
		& \colhead{(Jy/beam)} & \colhead{(\kms)}&
		\colhead{(mas $\times$  mas @ $\degr$)} } \startdata
	G033.64$-$00.22(M) & 18~53~32.56656 & $+$00~31~39.1152   &  ...  &...
	&  16.0 & 60.36 & 6.7 $\times$ 5.1 @ 165      \\
	J1848+0138     &18~48~21.81041 & $+$01~38~26.6296            & $-$1.3  & 1.1
	& 0.030 &       & 5.1 $\times$   3.6 @ 173      \\
	J1857$-$0048   &18~57~51.35887 & $-$00~48~21.9369          & 1.1 &  $-$1.3
	& 0.053 &       & 5.9 $\times$   3.0 @ 3      \\
	G035.57$-$00.03(W) & 18~56~22.52577 & $+$02~20~27.5007   &  ...  &...
	&  0.9 & 48.62 & 1.6 $\times$ 0.5 @ 164      \\
	J1855+0251     &18~55~35.43649 & $+$02~51~19.5650            & $-$0.2  & 0.5
	& 0.115 &       & 1.4 $\times$   0.5 @ 164      \\
	G041.15$-$00.20(M) & 19 07 14.36746 & $+$07 13 18.0190   &  ...  &...
	& 3.2 & 56.00 & 3.7 $\times$ 1.5 @ 4      \\
	J1905+0652     &19 05 21.21029 & $+$06 52 10.7830            & $-$0.5  & $-$0.4
	& 0.058 &       & 4.2 $\times$   2.4 @ 6     \\
	J1907+0907     &19 07 41.96333 & $+$09 07 12.3956            & 0.1  & 1.9
	& 0.072 &       & 3.9 $\times$   2.0 @ 179      \\
	G043.89$-$00.78(W) & 19 14 26.39610 & $+$09 22 36.5926   &  ...  &...
	&  8.6 & 59.18 & 1.4 $\times$ 0.4 @ 163      \\
	J1905+0952     &19 05 39.89897 & $+$09 52 08.4075            &  $-$2.2 & 0.5
	& 0.072 &       & 1.3 $\times$   0.6 @ 163      \\
	J1913+0932     &19 13 24.02535 & $+$09 32 45.3775            & $-$0.3  & 0.2
	& 0.047 &       & 1.5 $\times$   0.7 @ 159      \\
	J1922+0841     &19 22 18.63365 & $+$08 41 57.3753            &  1.9  & $-$0.7
	& 0.013 &       & 1.4 $\times$   0.6 @ 164      \\
	\enddata
	\tablecomments{Source names followed by M and W in parentheses indicate  methanol and water masers. $\Delta\Theta_{\rm E}$ and $\Delta\Theta_{\rm N}$ are the angular offsets of the QSOs relative to the maser toward the East and North. The absolute position, peak brightness, and (naturally weighted) beam size and position angle (PA; east of north) are listed for the first epoch. Note that the absolute position accuracies are limited to about $\pm 1$ mas from the assumed values for the QSOs. The local standard of rest (LSR) velocities ($V_{\rm LSR}$) of the reference maser spots are listed for \Ga, \Gc, and \Gd, whereas for \Gb, J1855+0251 was used as the phase reference.}
\end{deluxetable*}

We conducted parallax observations of two 6.7 GHz methanol (\meth) and two 22 GHz water (\hho) masers over 16 epochs spanning 1.2 yr with the National Radio Astronomy Observatory's (NRAO's)\footnote{NRAO is a facility of the National Science Foundation operated under cooperative agreement by Associated Universities, Inc.} Very Long Baseline Array (VLBA) under program BR210. 
{Prior to our program, G035.57$-$00.03 had not been observed for a parallax measurement.  Parallaxes for the other three sources had been previously published by \citet{2014Wu,2019ApJ...874...94W}. These were specifically chosen for re-observation, because previous parallax measurements (with only 4 or 5 epochs per source) were not accurate enough to provide strong constraints on spiral structure for distant portions of the Sagittarius arm.} 

Details of the observations are listed in Table \ref{table:observations}. We scheduled the observations to occur near the extrema of the sinusoidal parallax signatures in R.A., since the parallax amplitudes in decl. were considerably smaller.  
At each epoch, three 1.7-hr blocks of phase-referenced observations were inserted between four 0.5-hr geodetic blocks, used for clock and atmospheric delay calibrations \citep[see][for details]{2009ApJ...693..397R}. Four adjacent dual-polarized intermediate-frequency (IF) bands of 16 MHz bandwidth were used for the phase-referenced observations. The maser emissions were centered in the third IF.

The data were correlated in Socorro, New Mexico, with the
DiFX\footnote{DiFX, a software correlator for very long baseline interferometry (VLBI), is developed as part of the Australian Major National Research Facilities Programme by the Swinburne University of Technology and operated under licence.} software correlator \citep{2007PASP..119..318D}. The IF bands containing the masers were correlated with 4000 and 2000 spectral channels for the 6.7 GHz \meth\ and the 22 GHz \hho\ masers, yielding velocity channels of 0.18 and 0.11 \kms, respectively.

We reduced the correlated data with the Astronomical Image Processing
System \citep[AIPS;][]{2003ASSL..285..109G} and ParselTongue \citep{2006ASPC..351..497K} scripts in four steps as described in \cite{2009ApJ...693..397R}. In step 1, we corrected delays and phases for feed rotations, updated Earth's orientation parameters, ionospheric delays using total electron content maps, residual clock errors, tropospheric delays determined from the geodetic calibration blocks, and updated source positions (when necessary). In step 2, we converted correlator amplitudes to Jansky units by applying telescope gains and system temperatures.  In step 3, for the maser data we shifted frequencies to hold LSR velocities at a desired value for all sources and observations. In step 4, one scan on a strong calibrator was chosen to remove delay and phase differences among all bands. In step 5, for all but one source, a channel with strong and compact maser emission was used as the interferometer phase reference and applied to all maser channels and background quasi-stellar objects (QSOs). For \Gb, the bright background QSO (J1855+0251) was used as the phase reference, because the peak brightness of the H$_2$O maser (0.9 Jy beam$^{-1}$) provided insufficient signal-to-noise ratios on individual VLBA baselines in 8 kHz spectral channels and 20 s integrations.

The AIPS task IMAGR was used to image the spectral-line emission of the masers and continuum emission of the QSOs. The positions of the maser spots and background QSOs were determined by fitting elliptical Gaussian brightness distributions to the images using the AIPS task JMFIT. Table \ref{table:positions} lists the positions and brightnesses from the first epoch.

\begin{deluxetable*}{lrrrrrrr}[t]
	\tablecolumns{8} \tablewidth{0pc} \tablecaption{Astrometric Results \label{table:parallax}}
	\tablehead{ \colhead{Source} & \colhead{$\pi$} &  \colhead{$\mu_x$}
		& \colhead{$\mu_y$} & \colhead{\Vlsr} & \colhead{D$_{\pi}$} &\colhead{D$_{\rm k}$} & \colhead{D$_{\rm ave}$}  \\
		\colhead{} & \colhead{(mas)} &  \colhead{(\masy)}
		& \colhead{(\masy)} & \colhead{(\kms)}  & \colhead{(kpc)}  & \colhead{(kpc)}  & \colhead{(kpc)} } \startdata
\Ge                 & $0.076  \pm 0.011$  & $-3.11 \pm 0.16$ &$-6.36 \pm 0.17$ &    $ 41     \pm  3   $ &  $13.2^{+ 2.2}_{- 1.7}   $ & $12.5  \pm 0.8  $ &  $12.6 \pm 0.7$ \\
\Gf                 & $0.066  \pm 0.010$  & $-3.14 \pm 0.14$ &$-6.44 \pm 0.16$ &    $ 41     \pm  3   $ &  $15.2^{+ 2.7}_{- 2.0}   $ & $12.4  \pm 0.6  $ &  $12.7 \pm 0.6$ \\
G032.74$-$00.07$^{\ddagger}$     & $0.126  \pm 0.016$  & $-3.15 \pm 0.27$ &$-6.10 \pm 0.29$ &    $ 37     \pm  10  $ &  $7.9_{- 0.9}^{+ 1.2}    $ & $11.4  \pm 1.0  $ &  $10.6 \pm 0.8$ \\
\Ga$^{*\dagger \ddagger}$           & $0.103  \pm 0.011$  & $-3.01\pm0.07  $ &$-6.30 \pm 0.08$ &    $^{a}61  \pm  3   $ &  $9.7^{+ 1.2}_{- 1.0}   $ & $10.0  \pm 0.6  $ &  $9.9 \pm 0.5$ \\
\Gb$^{*}$           & $0.098  \pm 0.008$  & $-3.02 \pm 0.13$ &$-6.08 \pm 0.18$ &    $^{a}53  \pm  3   $ &  $10.2^{+ 0.9}_{- 0.8}   $ & $10.2  \pm 0.7  $ &  $10.2 \pm 0.6$ \\
G035.79$-$00.17     & $0.113  \pm 0.013$  & $-2.96 \pm 0.12$ &$-6.23 \pm 0.14$ &    $ 61     \pm  5   $ &  $8.8_{- 0.9}^{+ 1.2}    $ & $9.6   \pm  0.8 $ &  $9.4  \pm 0.6$ \\
G037.47$-$00.10     & $0.088  \pm 0.030$  & $-2.63 \pm 0.07$ &$-6.19 \pm 0.15$ &    $ 58     \pm  3   $ &  $11.4_{- 2.9}^{+ 5.9}   $ & $9.4   \pm  1.0 $ &  $9.6  \pm 0.9$ \\
G038.03$-$00.30     & $0.095  \pm 0.022$  & $-3.01 \pm 0.06$ &$-6.20 \pm 0.11$ &    $ 60     \pm  3   $ &  $10.5_{- 2.0}^{+ 3.2}   $ & $9.2   \pm  0.6 $ &  $9.3  \pm 0.6$ \\
\Gc$^{*\dagger}$           & $0.137  \pm 0.011$  & $-2.74 \pm 0.15$ &$-6.03 \pm 0.16$ &    $^{b}60  \pm  3   $ &  $7.3^{+ 0.6}_{- 0.5}    $ & $8.4   \pm  1.2 $ &  $7.6  \pm 0.5$ \\
G041.22$-$00.19     & $ 0.113 \pm 0.022$  & $-2.82 \pm 0.13$ &$-5.89 \pm 0.16$ &    $59      \pm  5   $ &  $8.8^{+ 2.1 }_{-  1.4 } $ & $8.5   \pm  1.4 $ &  $8.7  \pm 1.1$ \\
G043.03$-$00.45     & $ 0.130 \pm 0.019$  & $-3.03 \pm 0.15$ &$-6.56 \pm 0.20$ &    $56      \pm  5   $ &  $7.7^{+  1.3 }_{- 1.0 } $ & $8.3   \pm  0.7 $ &  $8.2  \pm 0.6$ \\
\Gd$^{*\dagger}$           & $ 0.136 \pm 0.005$  & $-3.01 \pm 0.16$ &$-6.03 \pm 0.18$ &    $^{c}54  \pm  3   $ &  $7.3^{+ 0.3}_{- 0.3}    $ & $8.3   \pm  0.8 $ &  $7.5  \pm 0.3$ \\
G045.07+00.13       & $ 0.129 \pm 0.007$  & $-3.21 \pm 0.26$ &$-6.11 \pm 0.26$ &    $59      \pm  5   $ &  $7.8^{+  0.4 }_{- 0.4 } $ & $7.7   \pm  1.0 $ &  $7.7  \pm 0.4$ \\
G045.45+00.06       & $ 0.119 \pm 0.017$  & $-2.34 \pm 0.38$ &$-6.00 \pm 0.54$ &    $55      \pm  7   $ &  $8.4^{+  1.4 }_{- 1.1 } $ & $7.6   \pm  1.4 $ &  $8.1  \pm 0.9$ \\
G045.49+00.12       & $ 0.144 \pm 0.024$  & $-2.62 \pm 0.17$ &$-5.61 \pm 0.16$ &    $58      \pm  3   $ &  $6.9^{+  1.4 }_{- 1.0 } $ & $7.0   \pm  1.6 $ &  $6.9  \pm 0.9$ \\
G045.80$-$00.35     & $ 0.137 \pm 0.023$  & $-2.52 \pm 0.17$ &$-6.08 \pm 0.27$ &    $64      \pm  5   $ &  $7.3^{+  1.5 }_{- 1.0 } $ & $6.3   \pm  1.5 $ &  $7.0  \pm 1.0$ \\
	\enddata
	\tablecomments{
 {Astrometric results are listed for the four sources discussed in this paper and 12 other sources in the Sagittarius arm at distances greater than about 8 kpc. }
 Column 1 lists Galactic source names.  Columns 2, 3, and 4 give parallaxes and proper motions in the eastward ($\mu_x$ =$\mu_\alpha\cos\delta$) and northward directions ($\mu_y$ = $\mu_\delta$ ).   Column 5 lists LSR velocities. Columns 6, 7, and 8 list the parallax distances, 3D kinematic distances, and their variance-weighted averages. We adopted these weighted averages in column 8 as the "final" distance.
{Sources with a superscript (*) are those reported in this paper. Their \Vlsr are determined from NH$_{3}(J, K) = (1, 1)$ emission and the references are (a) \citet{Wienen:2012}; (b) \citet{Pandian:2012}; and (c) \citet{Olmi:1993}.} For other sources parallaxes, proper motions, and line-of-sight velocities are taken from Table 1 of \citetalias{2019ApJ...885..131R} (along with their primary references), except for \Ge\ and \Gf, which are taken from     \cite{2021ApJS..253....1X}.
  Sources with a superscript ($\dagger$) are those having previously published parallaxes, and column 2 lists their weighted averages with the parallax measurements in this paper, as described in Section \ref{previous-parallax}.
  {Sources with a superscript ($\ddagger$) are those flagged as outliers in \citetalias{2019ApJ...885..131R}  but included in this paper when determining the characteristics of the Sagittarius arm, as described in Section \ref{sec-shape}.}
    }
\end{deluxetable*}

\section{Parallax and Distance Estimates} \label{sect-parallax}

\subsection{Parallax and Proper Motion Fitting}
We selected compact maser spots for parallax and proper motion fitting.
For the \hho\ masers \Gb\ and \Gd, we first did the parallax and proper 
motion fitting for each maser spot relative to each background source in order to identify and remove outliers caused by the blending of maser spots. We added ``error floors'' in the eastern and northern directions
in quadrature with the formal positional uncertainties from JMFIT and adjusted them to achieve a reduced $\chi^2$ per degree of freedom near unity in each coordinate.  The error floors were used to capture systematic errors in position measurements, usually owing to uncompensated atmospheric delays.

At the lower frequency of the 6.7 GHz methanol maser sources \Ga\ and \Gc, ionospheric ``wedges" can cause systematic positional shifts across the sky, significantly increasing astrometric error.  In order to mitigate this issue, we used an image-position-based method to generate positional data relative to an “artificial quasar” at the target maser position during each epoch. Detailed descriptions of this method can be found in \citet{2017AJ....154...63R} and \citet{2019AJ....157..200Z}.  We adopted this approach \citep{2019AJ....157..200Z} in the parallax fits.\footnote{A phase-calibration based procedure, MultiView~\citep{2017AJ....153..105R} and inverse MultiView~\citep{2022ApJ...932...52H}, has recently been shown superior to the image-based method, but could not be used with the observing strategy adopted in program BR210.}

Because all maser spots in a source should have the same parallax, we used all (unblended) maser spots and background sources to carry out a combined solution, which uses a single parallax parameter but allows each maser spot to have a different proper motion. The fitting results for the \hho\ and \meth\ masers are shown in Figures \ref{parallaxfit1} and \ref{parallaxfit2}, respectively. Tables \ref{table:detail} and \ref{table:ppmdetail} provide detailed results of the parallax and proper motion fits.  Since systematic errors caused by propagation delays are essentially the same for all maser spots, we multiplied the formal parallax uncertainties by the square root of the number of spots used for the parallax fitting \citep{2009ApJ...693..397R}. 

Individual fits for \Gd\ with different QSOs revealed some differences among the inferred error-floor values, suggesting different contributions to the systematic error budget from uncompensated atmospheric delays and/or variable unresolved QSO structure. After determining the individual error floors, we added these in quadrature to their formal uncertainties from JMFIT before doing a combined fit.  Also, as seen in Figure \ref{parallaxfit1}, the decl. data for \Gd\ show the presence of systematics in the residuals, which might be attributed to proper-motion acceleration, possibly owing to orbital acceleration in a long-period stellar binary.  Such effects have been seen in Very Long Baseline Interferometric data from Pleiades stars \citep{2014Sci...345.1029M}.  When adjustable acceleration parameters were included in the fitting process, these systematic residuals disappeared, as shown by the red dashed line in Figure \ref{parallaxfit1}.  { The estimated acceleration parameters are listed in Table \ref{table:detail}.  We adopted the parallax results from this approach for \Gd.}

The internal motions of maser features persisting for at least five epochs were averaged to estimate the motion of the central star.
Considering typical values of maser spot motions relative to the 
central star of $\sim5$ \kms\ \citep{2002ApJ...564..813M} for 
6.7 GHz \meth\ masers and $\sim10$ \kms\ \citep{1992ApJ...393..149G} 
for 22 GHz \hho\ masers, for sources with multiple maser features we inflated the formal uncertainties for each proper motion component of the central star by adding $\pm3$~\kms\ and $\pm5$~\kms\ ``error floors" in quadrature, respectively. 
For sources with only one maser feature, we adopted $\pm5$~\kms\ and $\pm10$~\kms\ error floors for the \meth\ and \hho\ masers. We adopted an LSR velocity for the central star based on the centroid of the NH$_{3}$ molecular line emission \citep{Wienen:2012,Pandian:2012,Olmi:1993}. 

\subsection{Previous Measurements}\label{previous-parallax}
{ 
Regarding \Ga, \citet{2014ApJ...783..130R}
reported a parallax of 0.131 $\pm$ 0.020 mas, whereas our data yield a parallax of 0.090 $\pm$ 0.014 mas. These two parallaxes are statistically consistent at the 1.7$\sigma$ level. Combining these two measurements with variance weighting yields a parallax of 0.103 $\pm$ 0.011 mas.

For \Gc, our data yield a parallax of 0.144 $\pm$ 0.014 mas, and combining this with the parallax of 0.125 $\pm$ 0.018 mas of \cite{2019ApJ...874...94W} yields a variance-weighted average parallax of 0.137 $\pm$ 0.011 mas.

For \Gd, parallaxes of 0.121 $\pm$ 0.020 mas and 0.144 $\pm$ 0.014 were reported by 
 \cite{2014Wu} and
\citet{2019ApJ...874...94W},
respectively. Combining these with the our measured parallax of 0.137 $\pm$ 0.006 mas yields a parallax of 0.136 $\pm$ 0.005  mas.
}

\subsection{3D Kinematic Distances}
The 3D kinematic distances provide an alternative method to estimate distance, independent of parallax measurements.  This technique combines likelihoods as a function of distance for line-of-sight velocities and proper motion components in Galactic longitude and latitude, assuming a rotation curve for the Galaxy.  Multiplying these three likelihoods yields a posteriori distance estimate, which is generally free of the two-fold ambiguities for standard (1D) kinematic distances using only line-of-sight velocities for sources within the Solar circle (ie, Galactocentric radii less than $R_0$). 

{
For the four sources discussed in this paper and 12 other sources in the Sagittarius arm  whose distances are greater than about 8 kpc, their 3D kinematic distances were estimated using the proper motions, LSR velocities and the Galactic rotation curve of \cite{2022AJ....164..133R}.}
{
Table \ref{table:parallax} lists their trigonometric parallaxes, proper motions, LSR velocities, the 3D kinematic distances, and a variance-weighted average 
\footnote{We first convert the measured parallax ($\pi \pm \sigma_\pi$) to distance ($d \pm \sigma_d$), where $d = 1/\pi$ and $\sigma_d = d^2 \sigma_\pi$, and then combine the parallax-converted distance and the 3D kinematic distance by variance weighting.} 
of the two distance estimates for each source. 
We adopted these average distances in the following discussion.} 

\begin{figure*}[t]
	\center
\includegraphics[scale=0.7]{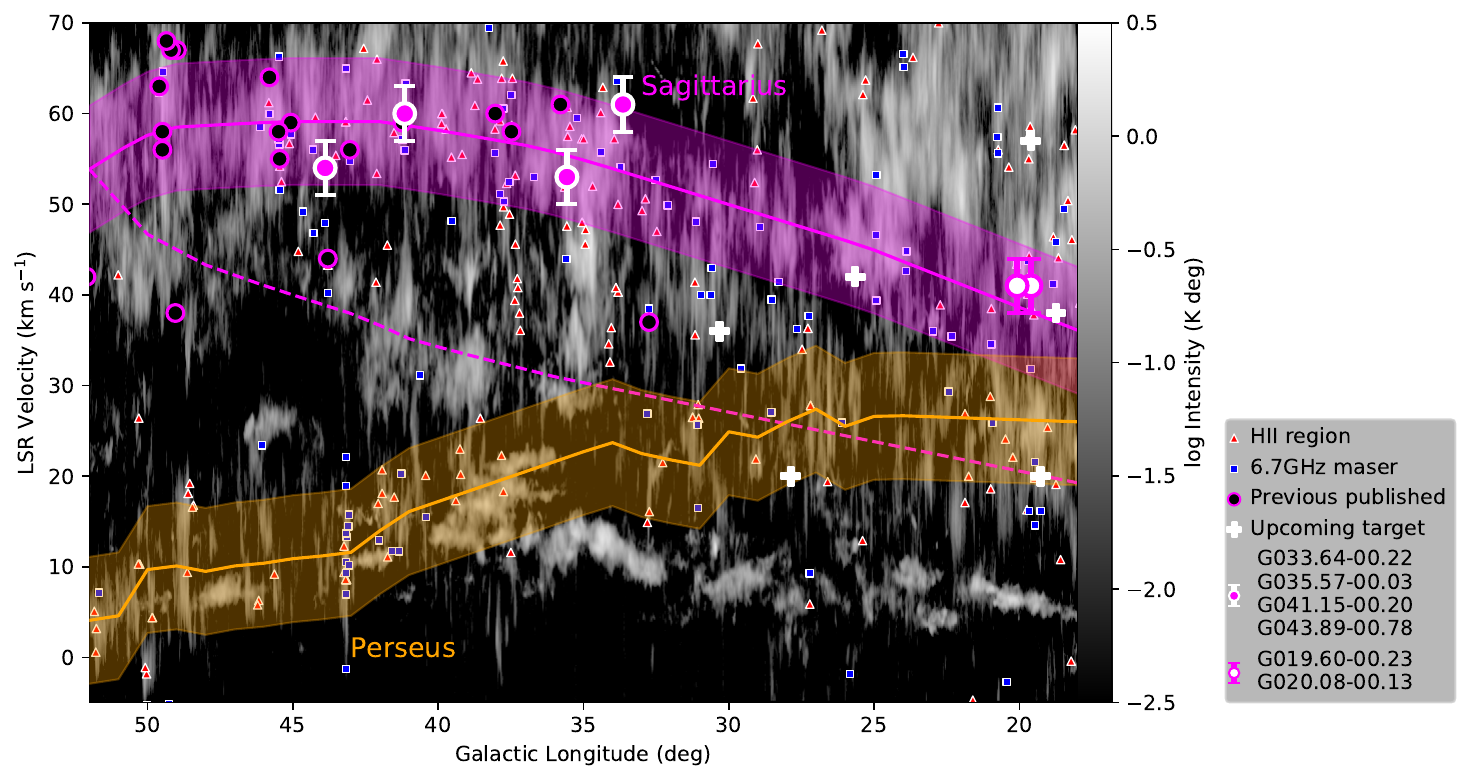}
	\caption{ Locations of sources superposed on a CO $(l, v)$ diagram from the Galactic Ring Survey \citep{2006ApJS..163..145J} integrated from b = $-1\deg$ to $+1\deg$. {Black dots with purple edges indicate previously published sources assigned to the far portion of the Sagittarius in the Table 1 of \citetalias{2019ApJ...885..131R} (along with their primary references). White crosses indicate the target for which we are performing VLBA parallax measurements under program BL312 as described in in Section \ref{sec-shape}. } Red  triangles and squares indicate the $(l, v)$ positions of \HII\ regions and 6.7 GHz \meth\ masers, estimated to be at  far kinematic distances in \dataset[WISE Catalog of Galactic \HII\ Regions]{http://astro.phys.wvu.edu/wise/}  \citep{2014ApJS..212....1A} and the GLOSTAR Galactic plane survey 6.7 GHz methanol maser catalogue \citep{2022A&A...666A..59N}, respectively.
 Traces of the Sagittarius and Perseus arms from \cite{2016ApJ...823...77R} pass through the CO \citep{2001ApJ...547..792D,2006ApJS..163..145J} and \HI\ \citep{2006AJ....132.1158S} emission features that define the arms in longitude, latitude, and velocity.
 The solid and dashed lines correspond to the far and near portions of the Sagittarius arm. The width of the shaded region corresponds to a $\pm$ 7 \kms velocity dispersion.
 \label{lv}
 }
\end{figure*}

\section{Discussion} \label{sect-discussion}
\subsection{Spiral Arm Assignments}\label{sec-location}

We now discuss the spiral arm assignments of the four sources from this paper and the two sources reported by \cite{2021ApJS..253....1X}. 
We assigned maser sources to the spiral arms based on their locations in H~\textsc{I} and/or CO longitude--velocity $(l, v)$ diagrams. 
{  
As shown in Figure \ref{lv}, 
traces of the Sagittarius and Perseus arms \citep{2016ApJ...823...77R} roughly follow 
the $^{13}$CO ($J$= 1--0) Galactic Ring Survey \citep{2006ApJS..163..145J} integrated from b = $-1\deg$ to $+1\deg$.
However, for Galactic longitudes lower than  about $30\deg$, the near portion of the Sagittarius arm overlaps with the velocities associated with the Perseus arm in $(l, v)$ plots.  And, for longitudes less than about $20\deg$, it is difficult to distinguish the far portion of the Sagittarius arm from the Perseus arm.    This highlights the need for accurate distance measurements to trace these arms in space.

The $(l, v)$ distribution of \HII\ regions and 6.7 GHz \meth\ masers, estimated to be at their far kinematic distances in \cite{2014ApJS..212....1A} and \cite{2022A&A...666A..59N}, reasonably well follow the Sagittarius and Perseus arms, at least down to Galactic longitudes of about $25\deg$.
}

\begin{figure*}[htbp]
	\center
	\includegraphics[scale=0.7]{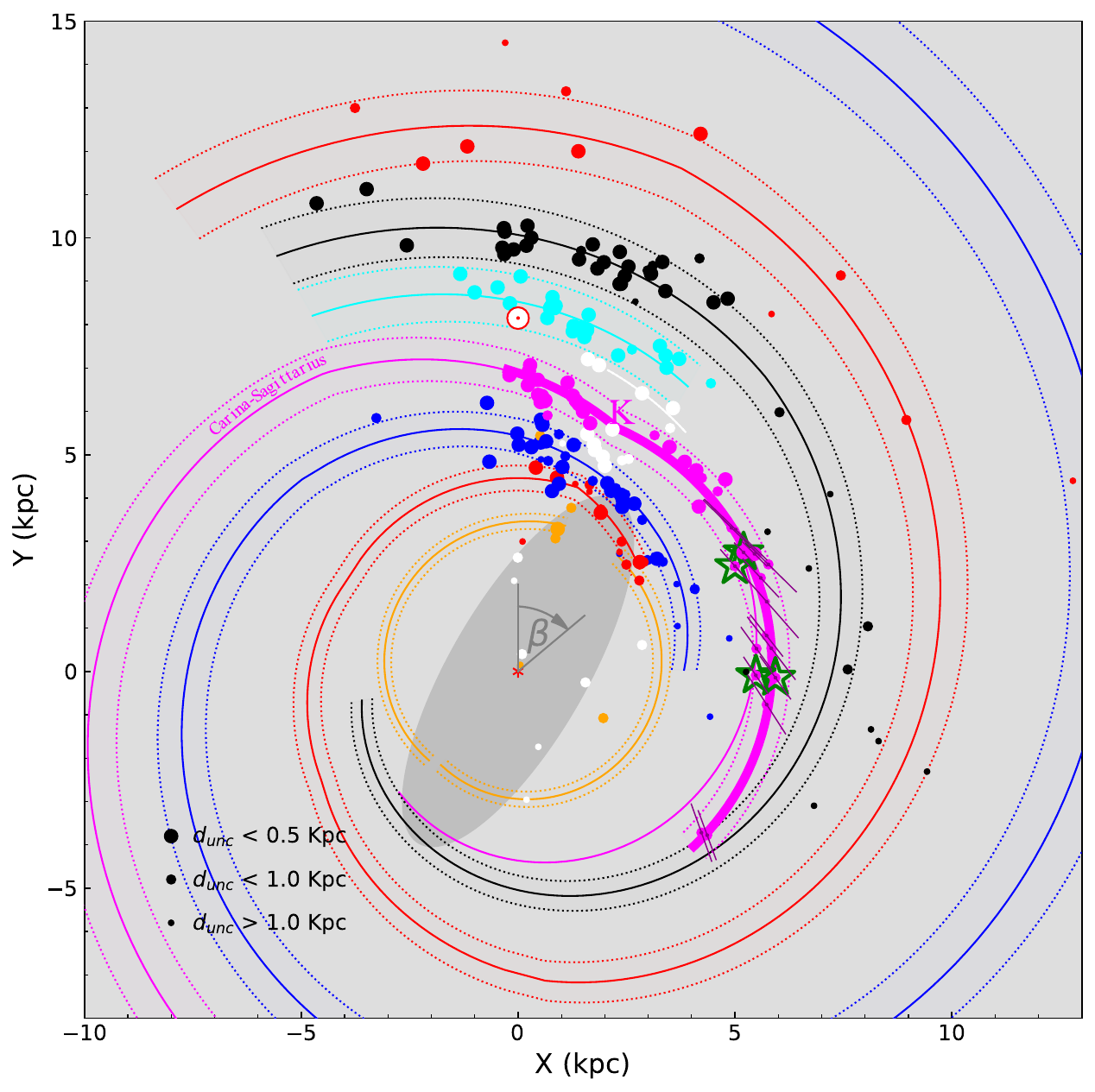}
	\caption{Plan view of the Milky Way as seen from the north Galactic pole following \citetalias{2019ApJ...885..131R}. The Galactic center is at (0,0) kpc and the Sun is at (0,8.15) kpc. Solid dots indicate parallax locations of the maser sources. Distance uncertainties are indicated by the inverse size of the symbols as given in the legend at the lower left.  Purple dots with 1$\sigma$ error bars indicate sources described in Section \ref{sect-parallax} and listed in Table \ref{table:parallax}.  Green stars indicate the locations of the sources reported in this paper.  The spiral arm models from \citetalias{2019ApJ...885..131R} are traced with solid lines and dotted lines enclose 90\% of the sources. The location of the kink in Sagittarius arm is marked with a "K."  The long Galactic bar is indicated with a shaded ellipse \citep{2015MNRAS.450.4050W}. The pitch angle fit for the Sagittarius arm determined in this paper is traced with a thick purple line.  \label{mk} }
\end{figure*}

As is evident in Figure \ref{lv}, the $(l, v)$ positions of \Ga, \Gb, \Gc, and \Gd\ indicate unambiguously an association with the far portion of the Sagittarius arm. Figure \ref{mk} shows the same four sources (green pentagrams) superposed on a map containing $\sim$200 maser sources listed in \citetalias{2019ApJ...885..131R}, as well as sources from \cite{2021ApJS..253....1X}, and \cite{2022AJ....163...54B}.  The locations of our four sources are also consistent with the far portion of the Sagittarius arm.  Therefore, we adopt that arm assignment for these four newly measured maser sources.

The distances for \Ge\ and \Gf\ are consistent with either the 
Sagittarius or Perseus arms.  
However, the far portion of the 
Sagittarius arm is favored over the Perseus arm by their $(l, v)$ 
positions, since, as shown in Figure \ref{lv}, their LSR velocities are
$\approx20$ \kms offset from those expected for the Perseus arm.
Typical LSR velocity deviations, owing to internal Virial motions, are $\approx7$ \kms\ from the average arm values \citep{2016ApJ...823...77R}. 
Also, at longitudes near 20\deg, the Galactic latitude of the far portion of the Sagittarius arm is $\approx-$0\d075, while that of the Perseus arm is $\approx+$0\d076 \citep{2016ApJ...823...77R,2019ApJ...885..131R}.  
At distances of over 12 kpc, the Galactic latitudes of \Ge\ ($-$0\d23) and \Gf\ ($-$0\d13) would place them more than 200 pc below the center of the Perseus arm.  Comparing this offset to the (Gaussian $1\sigma$) vertical dispersion of about 100 pc for the Perseus arm 
\citep[see Fig. 2 of][]{2016ApJ...823...77R} also favors association with the Sagittarius over the Perseus arm.  Considering all the evidence, we confidently assign \Ge\ and \Gf\ to the far portion of the Sagittarius arm.

\subsection{Structure of the Sagittarius Arm}\label{sec-shape}

\citetalias{2019ApJ...885..131R} fitted log-period spiral functions to the locations of HMSFRs in the Sagittarius arm with trigonometric parallax distances. They found that the Sagittarius arm has a ``kink" at Galactocentric azimuth $\beta \approx 24\deg$ with pitch angles 
of 17\d1 $\pm$ 1\d6 for $2\deg <\beta < 24\deg$, and 1\d0 $\pm$ 2\d1 for $ 24\deg< \beta < 97\deg$.  
Beyond the Galactic center (i.e., $\beta > 90\deg$), only two sources, G037.47$-$00.10 and G038.03$-$00.30, with parallaxes reported by \citet{2019ApJ...874...94W}, were used by \citetalias{2019ApJ...885..131R} to characterize the Sagittarius arm.  Our results for \Ga\ and \Gb\ together with the measurements of \Ge\ and \Gf\ \citep{2021ApJS..253....1X} add significant weight in determining the arm structure in this distant region.  
Using these six distant sources, adopting the averaged distances given in Table \ref{table:parallax}, and the same methodology as \citetalias{2019ApJ...885..131R}, we redetermine the characteristics of the Sagittarius arm over an extended range.  
{It is worth noting that, while employing the same procedure as \citetalias{2019ApJ...885..131R}, the increased accuracy of distances in Table \ref{table:parallax} has resulted in two \citetalias{2019ApJ...885..131R} outliers (G032.74-00.07 and G033.64-00.22, with $> 3\sigma$ residuals), now being within the acceptable range ($< 3\sigma$ residuals).
}

We estimate the pitch angles of the two Sagittarius arm segments to be 18\d6 $\pm$ 1\d4 and 1\d5 $\pm$ 1\d1 for segments between azimuths of $ -2\deg <\beta < 22\deg$ and $22\deg< \beta < 132\deg$, respectively. Our best-fitting parameters are consistent with the results of \citetalias{2019ApJ...885..131R} (17\d1 $\pm$ 1\d6 and 1\d0 $\pm$ 2\d1 for the segments between azimuths of $2\deg <\beta < 24\deg$ and $24\deg< \beta < 97\deg$, respectively).  Our results, based on more parallax data, extend the azimuth range from 97\deg to 132\deg\ (see Table \ref{table:pitchangles}). In addition, the significant decrease in the uncertainty of the pitch angle for $\beta > 22\deg$ between the estimate of \citetalias{2019ApJ...885..131R} and ours suggests that the pitch angle is nearly constant over the azimuth range $22\deg< \beta < 132\deg$.  We plot our extended model as the thick purple line in Figure \ref{mk}.

{While our fitted pitch angles are close to those of \citetalias{2019ApJ...885..131R}, our model for the Sagittarius arm in Fig. \ref{mk} starting near $\beta=42\deg$ traces radially outward from the \citetalias{2019ApJ...885..131R} model, and by $\beta=132\deg$ it is at $\approx1$ kpc greater radius.   Why does this occur? When the \citetalias{2019ApJ...885..131R} model was generated, \Ge\ and \Gf\ had not yet been measured, and G032.74$-$00.07 and G033.64$-$00.22 were flagged as outliers (and not used in their arm fitting).  Thus, the pitch angle for $\beta$ $>$ 42\deg\ was not well constrained.   In order to build a more complete spiral arm model for the Milky Way, for the distant portion of the Sagittarius arm they adopted a pitch angle of $8\deg$, which was closer to the average value determined for the spiral arms in the Milky Way and extrapolated the Sagittarius arm to the far end of the Galactic bar at approximately the same radius as that measured for the Norma arm at the near end of the bar.}

As seen in Figure \ref{mk}, our updated model for the far portion of the Sagittarius arm, based on new parallax and 3D kinematic distance measurements,
approaches the \citetalias{2019ApJ...885..131R} model for the Perseus arm
at { $\beta \approx 130\deg$}. This supports the suggestion \citep{2023ApJ...947...54X} that the Sagittarius and Perseus arm might merge.  However, we note that the location of the Perseus arm for azimuths greater than $\approx90\deg$ is currently not well constrained, and the Perseus and Sagittarius arms might merge closer to the far end of the bar, or they may not merge at all.  

{ As can be seen in the shaded regions of Figure \ref{lv}, there are dozens of 6.7 GHz masers at $\ell \approx$ 20\deg\ to 35\deg\ that could be associated with the far portion of the Sagittarius arm or Perseus arms. Parallax measurements for these sources using the MultiView calibration technique \citep{2017AJ....153..105R,2022ApJ...932...52H} could help us to test these possibilities.}
{In fact, for this purpose, under program BL312 (from March 2024 to March 2025), we are measuring the parallax of eight maser sources.  Six of the eight sources are in the scope of Figure \ref{lv}, shown as white crosses.}

\begin{deluxetable*}{lcrlrrrr}[tb]
	\tablecolumns{9} \tablewidth{0pc}
	\tablecaption{Sagittarius Arm Fitting Results \label{table:pitchangles} }
	\tablehead {
		\colhead{Reference} &$l$ Tangency &$\beta$ Range    &\colhead{$\beta_{\rm kink}$} &\colhead{$R_{\rm kink}$} &\colhead{$\psi_<$} &\colhead{$\psi_>$}&\colhead{Width}\\
		\colhead{}     &\colhead{(deg)}&\colhead{(deg)}    &\colhead{(deg)}          &\colhead{(kpc)}     &\colhead{(deg)}    &\colhead{(deg)}   &\colhead{(kpc)}
	}
	\startdata
	This paper                  & 284.4 &$-2\rightarrow132$ & $22\pm2$ &$6.06\pm0.06$ &$\p18.4\pm1.4$  &$\p1.7\pm1.0$  &$0.18\pm0.03$     \\
	\citetalias{2019ApJ...885..131R}  & 285.6 &$2\rightarrow\p97$& $24\pm2$ &$6.04\pm0.09$ &$\p17.1\pm1.6$  &$\p1.0\pm2.1$  &$0.27\pm0.04$     \\
	\enddata
	\tablecomments{Parameters estimated from fitting log-periodic spirals to the Sagittarius arm based on data from this paper, \cite{2021ApJS..253....1X}, and \citetalias{2019ApJ...885..131R}, assuming the distance to the Galactic center of $R_0$ = 8.15 kpc.  An arm tangency prior of $283\pm2\deg$  \citep{2000AA...358..521B} was used to constrain the fit. Column 2 lists the fitted tangency. Column 3 lists the azimuth range of the parallax data. Columns 4 and 5 list the Galactic azimuth and radius of the arm kink, separating two arm segments. Columns 6 and 7 give pitch angles for azimuths less and greater than $\beta_{\rm kink}$. Column 9 lists the intrinsic (Gaussian 1$\sigma$) arm width at $R_{\rm kink}$. Rows 1 and 2 give the best-fitting parameters in this paper and from \citetalias{2019ApJ...885..131R} for comparison. { }}
\end{deluxetable*}

\subsection{Kinematics of the Sagittarius Arm}
The peculiar motion components, \Us\ (toward the Galactic center), \Vs\ (in the direction of Galactic rotation), and \Ws\ (toward the north Galactic pole) {listed in Table \ref{table:3d} for the maser sources listed in Table \ref{table:parallax}} were calculated by adopting the distance to the Galactic center, $\Ro$ = 8.15 kpc; the circular rotation speed at the Sun, $\Theta_0$ = 236 \kms\; the solar motion values \U = 10.6 \kms, \V = 10.7 \kms, and \W = 7.6 \kms; and the Galactic rotation curve from \citetalias{2019ApJ...885..131R}.

\begin{figure}[t]
	\center
	\includegraphics[scale=0.62]{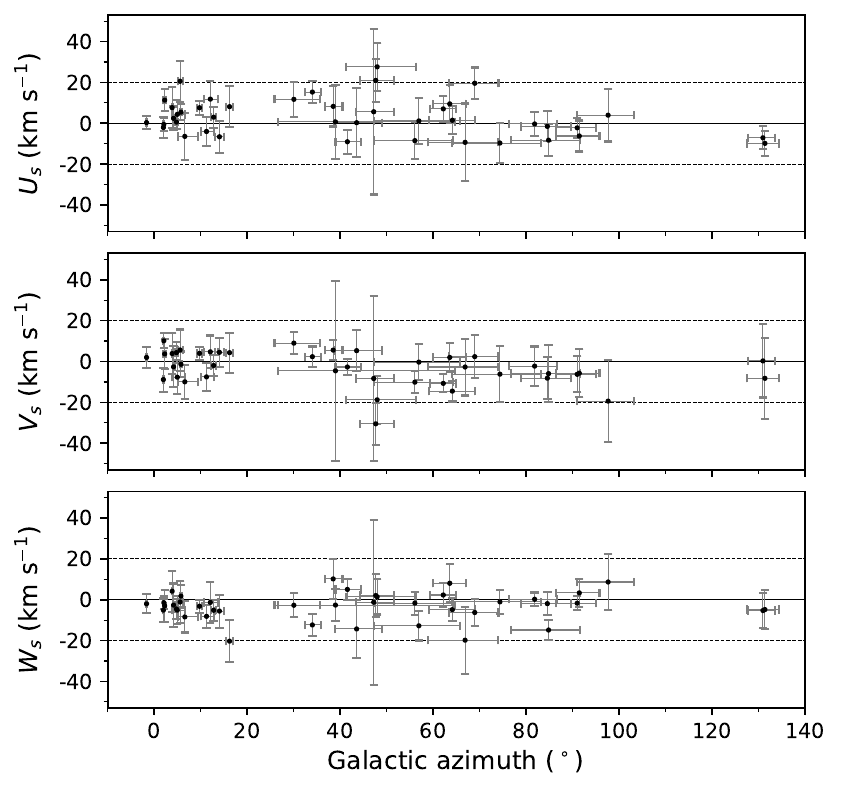}
	\caption{Plots of the peculiar motion components vs. Galactic azimuth {for sources in the Sagittarius arm based on the data from this paper and Table 1 of \citetalias{2019ApJ...885..131R}}. The top, middle, and bottom panels show \Us, \Vs, and \Ws, respectively, calculated by adopting model A5 and the Galactic rotation curve from \citetalias{2019ApJ...885..131R}.  \label{pm}}
\end{figure}

Figure \ref{pm} shows the peculiar motions components (\Us, \Vs, \Ws) as a function of Galactic azimuth ($\beta$) {for sources in the Sagittarius arm based on the data from this paper and Table 1 of \citetalias{2019ApJ...885..131R}}.  From Figure \ref{pm}, one can see that 
all values are consistent with 0 $\pm$ 20 \kms. The variance-weighted average peculiar motion components ($\overline{U_s}$, $\overline{V_s}$, $\overline{W_s}$) for the Sagittarius arm sources are listed in Table \ref{table:average3d}. 
For comparison with other spiral arms, the corresponding values for the sources in the Norma, Scutum, Local, Perseus, and Outer arms are also listed in Table \ref{table:average3d} (based on data in Table 1 of \citetalias{2019ApJ...885..131R}). Among these spiral arms, the Sagittarius arm has the among the smallest $\overline{U_s}$ and $\overline{V_s}$ magnitudes, indicating near circular Galactic orbits.  However, the Sagittarius arm has the only statistically significant $\overline{W_s}$ value, which comes from sources in the arm segment with $\beta<22\deg$ mostly moving toward the South Galactic Pole.

\begin{deluxetable}{lcr@{\phantom{1}$\pm$\phantom{1}}lr@{\phantom{1}$\pm$\phantom{1}}lr@{\phantom{1}$\pm$\phantom{1}}l}[t]
	\tablewidth{0pt} 
 \tablecaption{Average Peculiar Motions of HMSFRs in Spiral Arms \label{table:average3d}}
 	\tabletypesize{\scriptsize}
	\tablehead {\colhead{Arm} &
		\colhead{N}
		& \multicolumn{2}{c}{$\overline{U_s}$} & \multicolumn{2}{c}{$\overline{V_s}$} &
		\multicolumn{2}{c}{$\overline{W_s}$}
		\\
		\colhead{Name} & \colhead{} & \multicolumn{2}{c}{(\kms)}
		& \multicolumn{2}{c}{(\kms)} & \multicolumn{2}{c}{(\kms)} }
	\startdata
Norma  & 14  &  $   9.9$ & $   3.2$ & $  -7.7$ & $   1.7$ & $   0.2$ & $   1.7$ \\
Scutum  & 40  &  $  10.8$ & $   1.3$ & $  -1.4$ & $   1.1$ & $  -0.7$ & $   1.0$ \\
Sagittarius  & 42  & $   2.2$ & $   1.0$ & $  -0.8$ & $   1.0$ & $  -3.0$ & $   0.9$ \\
Local  & 28  &  $   0.1$ & $   1.0$ & $  -7.4$ & $   0.9$ & $   1.7$ & $   1.0$ \\
Perseus & 41  &  $   8.7$ & $   0.9$ & $  -5.9$ & $   1.1$ & $   0.1$ & $   1.0$ \\
Outer & 11  & $   5.7$ & $   1.9$ & $  -6.7$ & $   2.6$ & $   0.7$ & $   2.4$ \\
	\enddata
	\tablecomments {Variance-weighted averages of peculiar motions were calculated from data in this paper and Table 1 of \citetalias{2019ApJ...885..131R}. Columns 1 and 2 list the arm name and the number of sources. Columns 3, 4, and 5 list the variance-weighted average of the peculiar motion components \Us, \Vs, and \Ws.} 
\end{deluxetable}

\section{Summary} \label{sec-summary}

We measured the parallaxes and proper motions of four masers in HMSFRs
associated with the distant portions of the Sagittarius spiral arm. The results for 
\Gc\ and \Gd\ at Galactic azimuth $\beta\sim80\deg$ are consistent with the previous
model for the arm by \citetalias{2019ApJ...885..131R}.  However, the more distant sources,
\Ga\ and \Gb, as well as \Ge\ and \Gf\ from \cite{2021ApJS..253....1X}, better constrain the structure of the Sagittarius arm beyond the Galactic center
($\beta > 90\deg$) and suggest that the Sagittarius arm model of \citetalias{2019ApJ...885..131R} should be moved outward by about 1 kpc
at {$\beta\approx130\deg$}, where it might merge with the Perseus arm.

\begin{acknowledgements}
This work was funded by the National SKA Program of China (grant No. 2022SKA0120103), NSFC grant 11933011, 12303072,  the Key Laboratory for Radio Astronomy, the Jiangsu Funding Program for Excellent
Postdoctoral Talent (grant No. 2023ZB093), and the Natural Science Foundation of Jiangsu Province (grant No. BK20210999).
\end{acknowledgements}

\vskip 0.5truein
\facility{VLBA}

\clearpage

\clearpage
\appendix

\restartappendixnumbering

\section{Observations}
Here, we list the details of the epochs observed in Table \ref{table:observations}.

\begin{table*}[tb]
	\centering
	\caption{Dates of VLBA Observations}
	\label{table:observations}
	\begin{tabular}{cccc}
	\toprule
    \Ga & \Gb & \Gc & \Gd  \\
	\hline
	2015 Mar 01 & 2015 Mar 08 & 2015 Mar 01  & 2015 Mar 08   \\
	2015 Mar 28 & 2015 Mar 27 & 2015 Mar 28  & 2015 Mar 27   \\
	2015 Apr 22 & 2015 Apr 20 & 2015 Apr 22  & 2015 Apr 20   \\
	2015 May 20 & 2015 May 14 & 2015 May 20  & 2015 May 14   \\
	2015 Aug 28 & 2015 Aug 29 & 2015 Aug 28  & 2015 Aug 29   \\
	2015 Sep 10 & 2015 Sep 13 & 2015 Sep 10  & 2015 Sep 13   \\
	2015 Sep 21 & 2015 Sep 28 & 2015 Sep 21  & 2015 Sep 28   \\
	2015 Oct 03 & 2015 Oct 09 & 2015 Oct 03  & 2015 Oct 09   \\
	2015 Oct 16 & 2015 Oct 24 & 2015 Oct 16  & 2015 Oct 24   \\
	2015 Oct 27 & 2015 Nov 02 & 2015 Oct 27  & 2015 Nov 02   \\
	2015 Nov 06 & 2015 Nov 13 & 2015 Nov 06  & 2015 Nov 13   \\
	2015 Nov 16 & 2015 Nov 23 & 2015 Nov 16  & 2015 Nov 23   \\
	2016 Feb 26 & 2016 Feb 25 & 2016 Feb 26  & 2016 Feb 25   \\
	2016 Mar 25 & 2016 Mar 22 & 2016 Mar 25  & 2016 Mar 22   \\
	2016 Apr 18 & 2016 Apr 16 & 2016 Apr 18  & 2016 Apr 16   \\
	2016 May 26 & 2016 May 25 & 2016 May 26  & 2016 May 25   \\
	\bottomrule
\end{tabular}
\end{table*}

\section{Parallax and Proper Motion Fits}
Here, we list the details of the parallaxes fits with formal uncertainties and the proper motion estimations listed in Table 
\ref{table:detail} and Table \ref{table:ppmdetail}, respectively.

\begin{figure*}[tb]
	\begin{minipage}[c]{0.5\linewidth}
\includegraphics[scale=0.5,trim=0 0 0 0,clip]{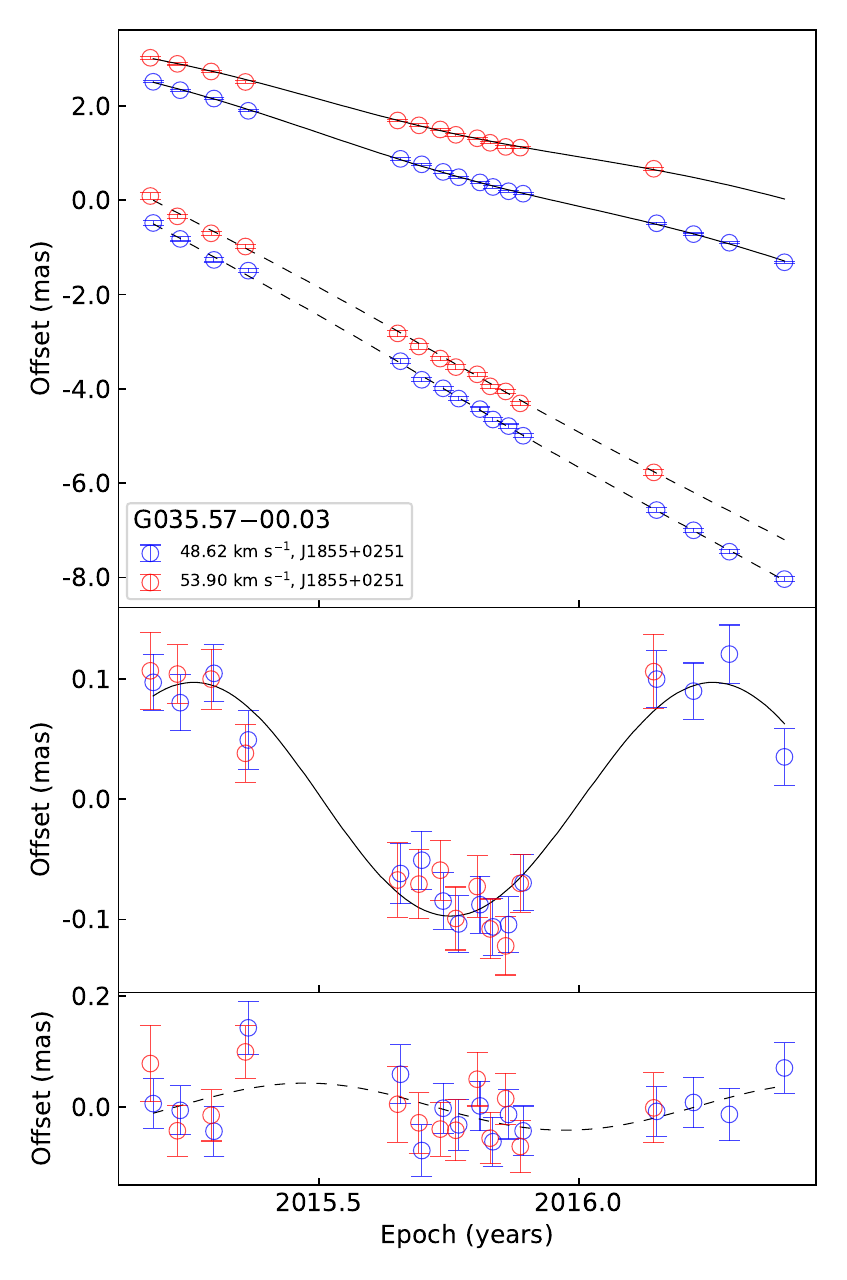}
\centering
	\end{minipage}
	\hfill
	\begin{minipage}[c]{0.5\linewidth}
\includegraphics[scale=0.5,trim= 0 0  0 0,clip]{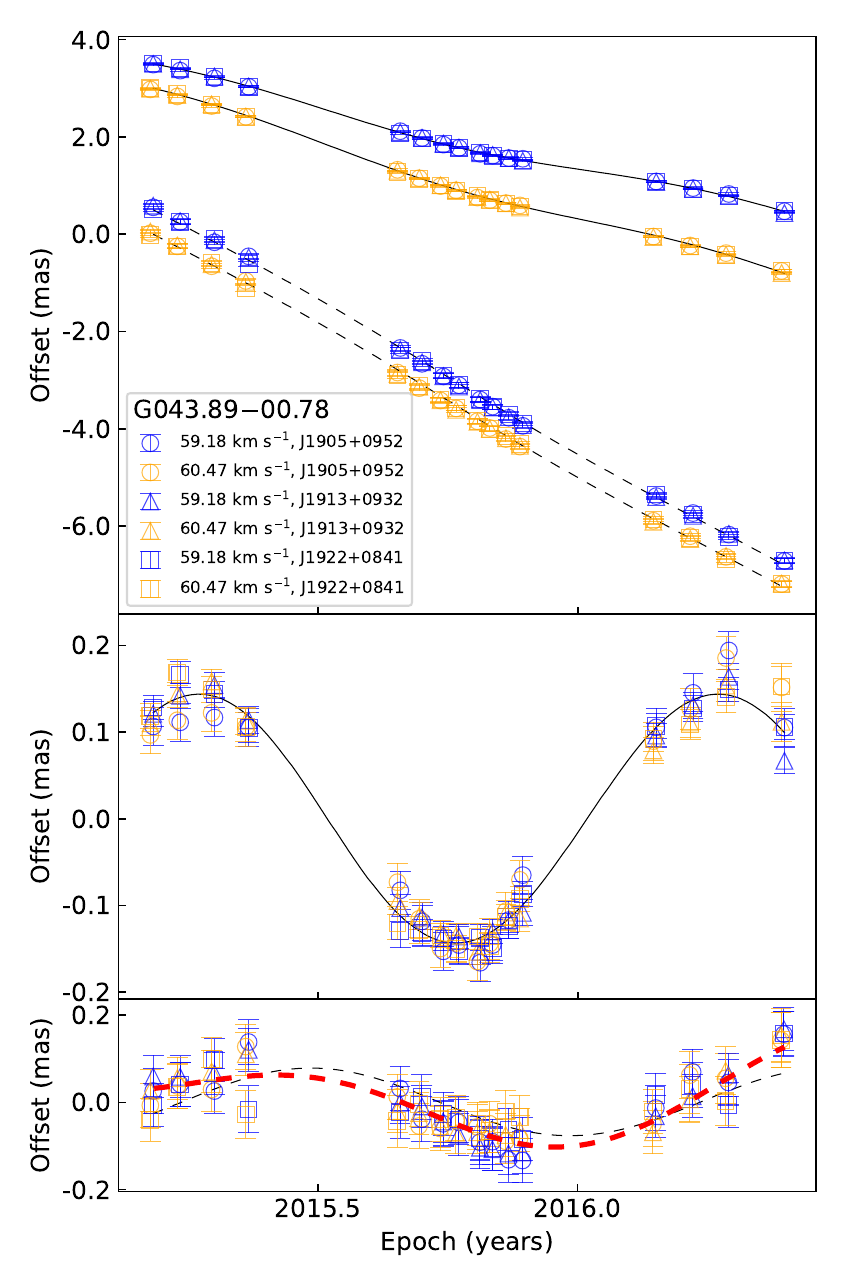}
	\end{minipage}
	\caption{Parallax fitting results for the \hho\ masers \Gb\ and \Gd. The top panels give the eastward (solid line) and northward (dashed line) position offsets vs. time. The middle and bottom panels display the eastward and northward data with the fitted proper motion removed. See the legend for the source names, the \Vlsr of each maser spot, and the background QSO(s) used for the position reference(s). {  The red dashed line shows the northward position offsets when proper-motion acceleration was included in the fitting process for \Gd. }
 \label{parallaxfit1}}
\end{figure*}

\begin{figure*}[tb]
\begin{minipage}[c]{0.5\linewidth}
\centering
\includegraphics[scale=0.5,trim=0 0 0 0,clip]{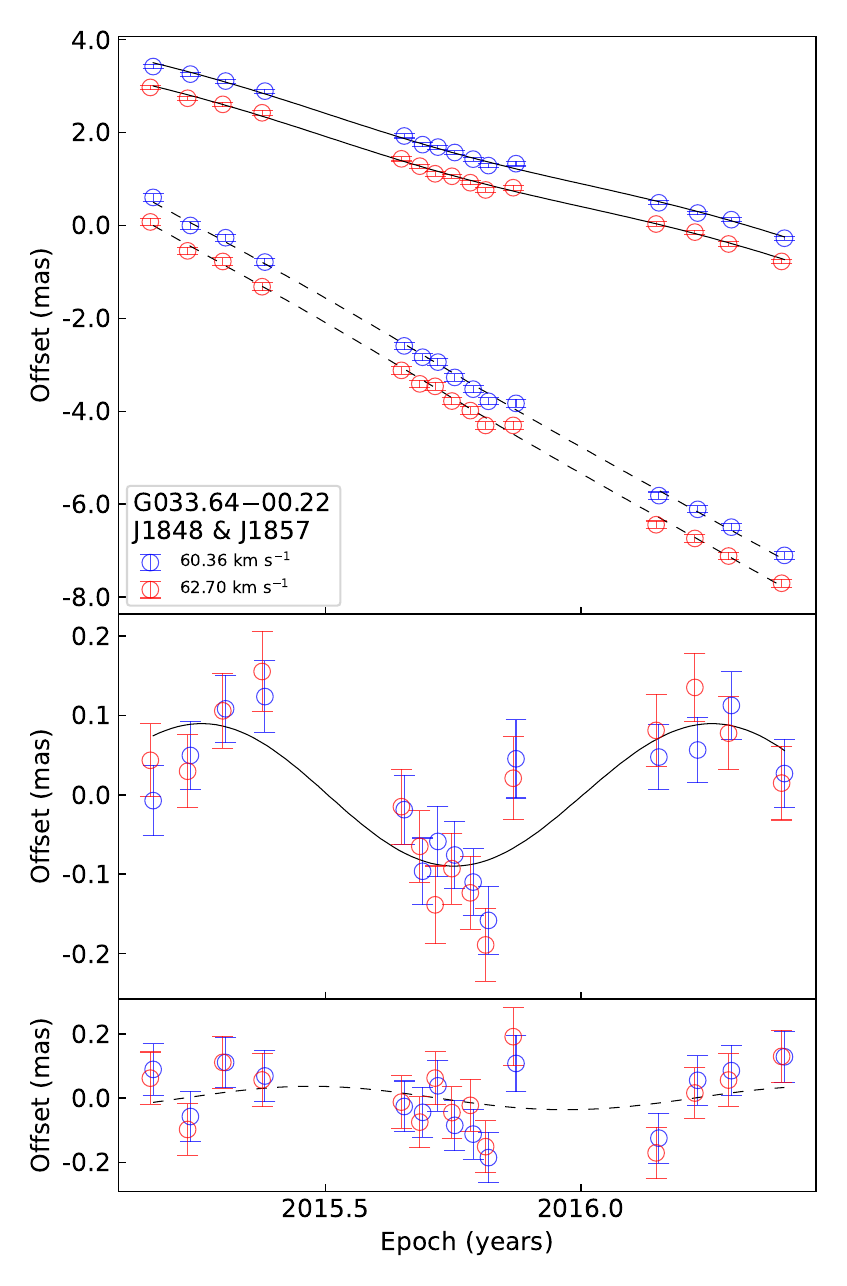}
	\end{minipage}
	\hfill
	\begin{minipage}[c]{0.5\linewidth}
\includegraphics[scale=0.5,trim=0 0 0 0,clip]{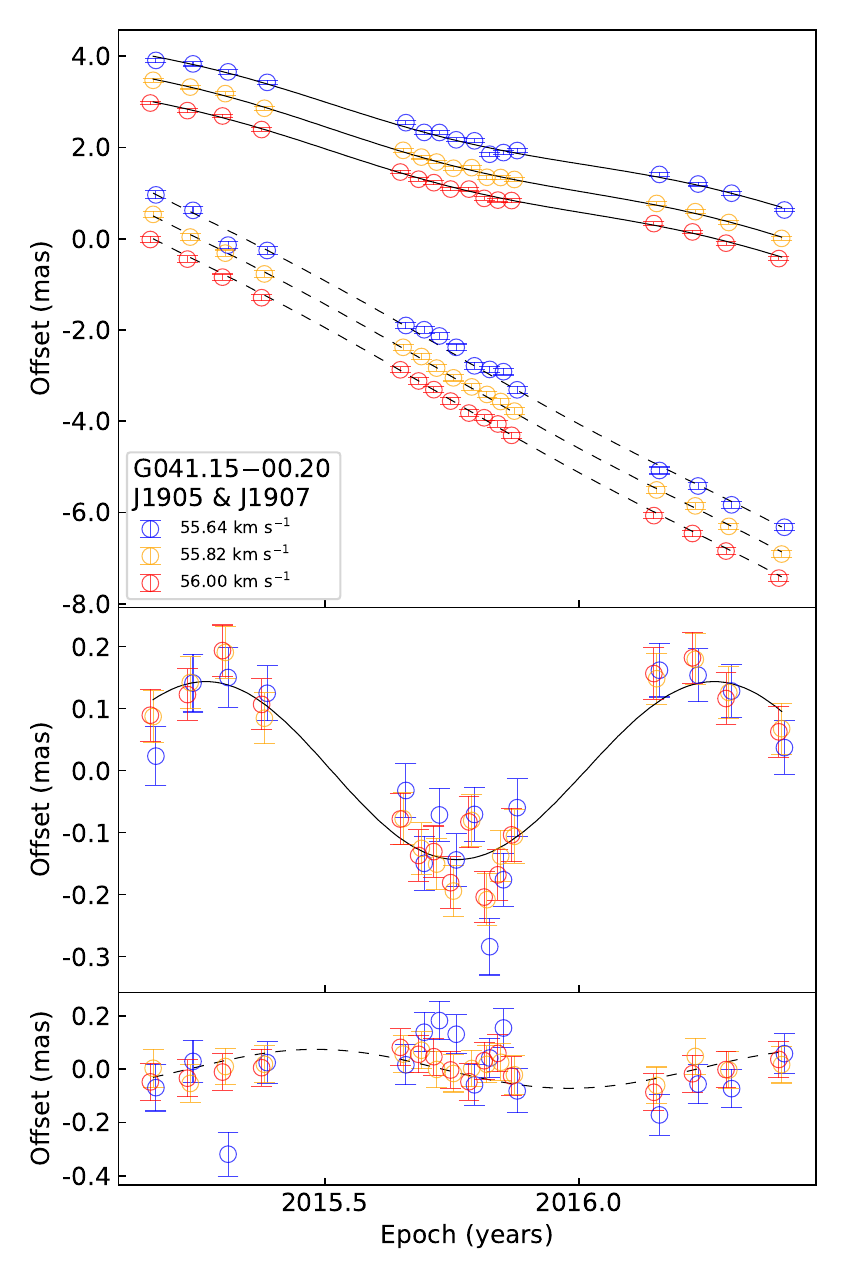}
	\end{minipage}
	\caption{Same as Figure \ref{parallaxfit1}, but for the \meth\ masers, \Ga\ and \Gc. Their parallaxes are estimated using the approach of \citet{2019AJ....157..200Z}.	\label{parallaxfit2}}
\end{figure*}


\begin{deluxetable*}{lcccccrr}[tb]
\tablecaption{Detailed Results of the Parallaxes \label{table:detail} }
\tablehead{\colhead{Target} & \colhead{Background} & \colhead{\Vlsr} & \colhead{Detected}
    & \colhead{Parallax} & \colhead{Solved for } & \colhead{acc$_x$} & \colhead{acc$_y$} \\
    \colhead{Source} & \colhead{Source} & \colhead{(\kms)} & \colhead{epochs} & \colhead{(mas)} & \colhead{acceleration} &\colhead{(mas~y$^{-2}$)} &\colhead{(mas~y$^{-2}$)}
    }
\startdata
\multicolumn{5}{c}{\hho\ maser} \\
\hline
\Gb & J1855+0251  & 48.62 & 1111~1111 1111~1111 &     0.098 $\pm$ 0.006  &No&&  \\
&             & 53.90 & 1111~1111 1111~1000 &    0.098 $\pm$ 0.013   &No&&  \\
&\multicolumn{2}{c}{Combined fit} && 0.098 $\pm$ 0.008  &No&&   \\
\hline
\Gd & J1905+0952& 59.18 & 1111 11111111 1111 &     0.146 $\pm$ 0.006     &No&& \\
&           & 60.47 & 1111 11111111 1111 &     0.142 $\pm$ 0.006     &No&&  \\
& J1913+0932& 59.18 & 1111 11111111 1111 &     0.143 $\pm$ 0.004     &No&& \\
&           & 60.47 & 1111 11111111 1111 &     0.140 $\pm$ 0.004     &No&& \\
& J1922+0841& 59.18 & 1111 11111111 1111 &     0.150 $\pm$ 0.003     &No&& \\
&           & 60.47 & 1111 11111111 1111 &     0.145 $\pm$ 0.005     &No&& \\
&\multicolumn{2}{c}{Combined fit} & &0.145 $\pm$ 0.003 &No&& \\
\hline
\Gd & J1905+0952& 59.18 & 1111 11111111 1111 &     0.153 $\pm$   0.012     & Yes & $-$0.12 $\pm$    0.16 & 0.62 $\pm$    0.12 \\
&           & 60.47 & 1111 11111111 1111 &     0.129 $\pm$   0.013     & Yes &    0.16 $\pm$    0.18 & 0.48 $\pm$    0.12  \\
& J1913+0932& 59.18 & 1111 11111111 1111 &     0.157 $\pm$   0.007     & Yes & $-$0.21 $\pm$    0.09 & 0.73 $\pm$    0.05 \\
&           & 60.47 & 1111 11111111 1111 &     0.130 $\pm$   0.009     & Yes &    0.11 $\pm$    0.13 & 0.55 $\pm$    0.09 \\
& J1922+0841& 59.18 & 1111 11111111 1111 &     0.140 $\pm$   0.007     & Yes &    0.11 $\pm$    0.09 & 0.43 $\pm$    0.15 \\
&           & 60.47 & 1111 11111111 1111 &     0.119 $\pm$   0.010     & Yes &    0.36 $\pm$    0.14 & 0.26 $\pm$    0.16 \\
&\multicolumn{2}{c}{Combined fit} & &0.137 $\pm$ 0.006   & Yes &    0.08 $\pm$    0.06 & 0.54 $\pm$    0.05  \\
\hline
\hline
    \multicolumn{5}{c}{\meth\ maser} \\
\hline
\Ga & J848 \& J1857 & 60.36& 1111~1111 1101~1111 &  0.081 $\pm$ 0.018   &No&&  \\
&                & 62.70& 1111~1111 1101~1111 &  0.099 $\pm$ 0.019   &No&&  \\
\multicolumn{2}{c}{Combined fit} & && 0.090 $\pm$ 0.014   &No&&  \\
\hline
\Gc & J1905 \& J1907 & 55.64 &1111~1111 1111~1111 &  0.128 $\pm$ 0.026   &No&&  \\
&                & 55.82 &1111~1111 1111~1111 &  0.140 $\pm$ 0.016   &No&&  \\
&                & 56.00 &1111~1111 1111~1111 &  0.141 $\pm$ 0.015   &No&&  \\
\multicolumn{2}{c}{Combined fit} & & & 0.144 $\pm$ 0.014  &No&&   \\
\enddata
\end{deluxetable*}

\begin{deluxetable*}{lcccrrr}[tb]
\tablecaption{Detailed Results of the Proper Motions \label{table:ppmdetail} }
\tablehead{\colhead{Target}  & \colhead{Feature} & \colhead{\Vlsr}
& \colhead{$\mu_x$ }& \colhead{$\mu_y$}& \colhead{$\Delta x$}& \colhead{$\Delta y$}\\
	\colhead{Source} & \colhead{} & \colhead{(\kms)}  & \colhead{(\masy)} & \colhead{(\masy)} & \colhead{(mas)}& \colhead{(mas)} }
\startdata
		\multicolumn{7}{c}{\hho\ maser} \\		\hline
		\Gb  & 1 & 48.19$\sim$49.05 &   $-$3.11 $\pm$ 0.04 &   $-$6.26 $\pm$ 0.07 &    0 &    0 \\
     & 2 & 64.50$\sim$68.71 &   $-$3.23 $\pm$ 0.11 &   $-$6.02 $\pm$ 0.21 &   624 & $-$1088 \\
     & 3 & 55.75$\sim$56.07 &   $-$3.17 $\pm$ 0.07 &   $-$6.16 $\pm$ 0.17 &  $-$319 &  $-$303\\
     & 4 & 50.56$\sim$54.88 &   $-$2.45 $\pm$ 0.06 &   $-$5.93 $\pm$ 0.11 &    53 &    46 \\
     & 5 & 53.16$\sim$53.59 &   $-$3.16 $\pm$ 0.12 &   $-$6.04 $\pm$ 0.20 &  1428 &   596 \\
     &\multicolumn{2}{c}{Average}       & $-$3.02 $\pm$ 0.08    & $-$6.08 $\pm$ 0.15  & &   \\
\multicolumn{3}{c}{Enlarge 5 \kms\ Error} & $-$3.02 $\pm$ 0.13    & $-$6.08 $\pm$ 0.18  & &   \\
     \hline
     \Gd  & 1 & 58.10$\sim$59.61 & $-$2.48 $\pm$ 0.03	& $-$6.08 $\pm$ 0.03  & 0  & 0 \\
     & 2 & 48.71$\sim$51.73 & $-$3.42 $\pm$ 0.04	& $-$5.02 $\pm$ 0.05  & $-$62 &   $-$51  \\
     & 3 & 59.72$\sim$60.80& $-$3.09 $\pm$ 0.03	& $-$5.81 $\pm$ 0.13  & $-$78 &   $-$102  \\
     & 4 & 54.43$\sim$56.37 & $-$3.03 $\pm$ 0.10	& $-$7.22 $\pm$ 0.14  & $-$65 &   $-$276  \\
     &\multicolumn{2}{c}{Average}       & $-$3.01 $\pm$ 0.05    & $-$6.03 $\pm$ 0.09  & &   \\
\multicolumn{3}{c}{Enlarge 5 \kms\ Error} & $-$3.01 $\pm$ 0.16    & $-$6.03 $\pm$ 0.18  & &   \\		\hline		\hline
			\multicolumn{7}{c}{\meth\ maser} \\		\hline
\Ga
& 1 & 60.18$\sim$60.54 &   $-$3.01 $\pm$ 0.03 &   $-$6.26 $\pm$ 0.05 &    0 &    0 \\
& 2 & 62.52$\sim$63.24 &   $-$3.00 $\pm$ 0.04 &   $-$6.33 $\pm$ 0.06 & $-$28 &  $-$2\\
&\multicolumn{2}{c}{Average}       & $-$3.01 $\pm$ 0.04    & $-$6.30 $\pm$ 0.06  & &   \\
\multicolumn{3}{c}{Enlarge 3 \kms\ Error} & $-$3.01 $\pm$ 0.07    & $-$6.30 $\pm$ 0.08  & &   \\
\hline
\Gc
& 1 & 55.65$\sim$56.00 &   $-$2.74 $\pm$ 0.03 &   $-$6.03 $\pm$ 0.06 &    0 &    0 \\
\multicolumn{3}{c}{Enlarge 5 \kms\ Error} & $-$2.74 $\pm$ 0.15    & $-$6.03 $\pm$ 0.16  & &   \\
\hline
\enddata
\end{deluxetable*}

\section{Peculiar Motions}
Here, we list the peculiar motions of the sources located in the Sagittarius arm in Table \ref{table:3d}.

\begin{deluxetable*}{lcr@{\phantom{.} $\pm$ \phantom{.}}lr@{\phantom{.} $\pm$ }lr@{\phantom{.} $\pm$ \phantom{.}}lcr@{\phantom{.} $\pm$ \phantom{.}}lr@{\phantom{.} $\pm$ \phantom{.}}lr@{\phantom{.} $\pm$ \phantom{.}}l}[tb]
\tabletypesize{\scriptsize}
\tablewidth{10pt} \tablecaption{Peculiar Motions \label{table:3d}}
\tablehead {
	\colhead{Source} & \colhead{$\beta$}
	& \multicolumn{2}{c}{$U_{s}$} & \multicolumn{2}{c}{$V_{s}$} &
	\multicolumn{2}{c}{$W_{s}$} & \colhead{Distance}& \multicolumn{2}{c}{$\mu_{x}$}& \multicolumn{2}{c}{$\mu_{y}$}& \multicolumn{2}{c}{\vlsr} \\
	\colhead{}   & \colhead{($\deg$)}   & \multicolumn{2}{c}{(\kms)}
	& \multicolumn{2}{c}{(\kms)} & \multicolumn{2}{c}{(\kms)} & \colhead{kpc}& \multicolumn{2}{c}{\masy}& \multicolumn{2}{c}{\masy}& \multicolumn{2}{c}{\kms}  }
\startdata
 G045.49$+$00.12 & $  56.0_{-  8.7}^{+  8.2}$ & $  -8.5$ & $   9.2$ & $ -10.1$ & $   5.4$ & $  -1.6$ & $   5.6$  & $ 6.9^{+0.9}_{- 0.9}$ & $-2.62$ & $ 0.17$ & $-5.61$ & $ 0.16$ & $   58$ & $    3$  \\
 G045.80$-$00.35 & $  56.9_{-  9.6}^{+  8.9}$ & $   1.1$ & $  11.3$ & $  -0.3$ & $   9.0$ & $ -12.8$ & $   7.2$  & $ 7.0^{+1.0}_{- 1.0}$ & $-2.52$ & $ 0.17$ & $-6.08$ & $ 0.27$ & $   64$ & $    5$  \\
 G043.89$-$00.78 & $  62.2_{-  2.8}^{+  2.8}$ & $   7.0$ & $   6.3$ & $ -10.7$ & $   4.4$ & $   2.3$ & $   5.9$  & $ 7.5^{+0.3}_{- 0.3}$ & $-3.01$ & $ 0.16$ & $-6.03$ & $ 0.18$ & $   54$ & $    3$  \\
 G045.07$+$00.13 & $  63.6_{-  3.6}^{+  3.5}$ & $   9.5$ & $   9.4$ & $   2.0$ & $   7.3$ & $   8.1$ & $   9.4$  & $ 7.7^{+0.4}_{- 0.4}$ & $-3.21$ & $ 0.26$ & $-6.11$ & $ 0.26$ & $   59$ & $    5$  \\
 G041.15$-$00.20 & $  64.1_{-  5.1}^{+  4.8}$ & $   1.5$ & $   6.8$ & $ -14.5$ & $   5.1$ & $  -4.9$ & $   5.5$  & $ 7.6^{+0.5}_{- 0.5}$ & $-2.74$ & $ 0.15$ & $-6.03$ & $ 0.16$ & $   60$ & $    3$  \\
 G045.45$+$00.06 & $  66.9_{-  8.0}^{+  7.2}$ & $  -9.3$ & $  19.0$ & $  -2.7$ & $  13.9$ & $ -19.8$ & $  16.4$  & $ 8.1^{+0.9}_{- 0.9}$ & $-2.34$ & $ 0.38$ & $-6.00$ & $ 0.54$ & $   55$ & $    7$  \\
 G043.03$-$00.45 & $  68.9_{-  5.5}^{+  5.1}$ & $  19.6$ & $   7.7$ & $   2.4$ & $  10.6$ & $  -6.2$ & $   6.4$  & $ 8.2^{+0.6}_{- 0.6}$ & $-3.03$ & $ 0.15$ & $-6.56$ & $ 0.20$ & $   56$ & $    5$  \\
 G041.22$-$00.19 & $  74.3_{- 10.3}^{+  8.8}$ & $  -9.7$ & $   9.7$ & $  -6.2$ & $  13.8$ & $  -1.1$ & $   5.8$  & $ 8.7^{+1.1}_{- 1.1}$ & $-2.82$ & $ 0.13$ & $-5.89$ & $ 0.16$ & $   59$ & $    5$  \\
 G038.03$-$00.30 & $  81.8_{-  5.4}^{+  4.9}$ & $  -0.3$ & $   5.8$ & $  -2.3$ & $   9.6$ & $   0.2$ & $   3.3$  & $ 9.3^{+0.6}_{- 0.6}$ & $-3.01$ & $ 0.06$ & $-6.20$ & $ 0.11$ & $   60$ & $    3$  \\
 G035.79$-$00.17 & $  84.5_{-  5.7}^{+  5.1}$ & $  -1.5$ & $   7.3$ & $  -8.2$ & $  10.2$ & $  -1.9$ & $   5.6$  & $ 9.4^{+0.6}_{- 0.6}$ & $-2.96$ & $ 0.12$ & $-6.23$ & $ 0.14$ & $   61$ & $    5$  \\
 G037.47$-$00.10 & $  84.8_{-  8.0}^{+  6.8}$ & $  -8.3$ & $   7.9$ & $  -5.9$ & $  14.1$ & $ -14.8$ & $   4.7$  & $ 9.6^{+0.9}_{- 0.9}$ & $-2.63$ & $ 0.07$ & $-6.19$ & $ 0.15$ & $   58$ & $    3$  \\
 G033.64$-$00.22 & $  91.0_{-  4.5}^{+  4.1}$ & $  -2.2$ & $   4.9$ & $  -6.3$ & $   8.8$ & $  -1.7$ & $   3.4$  & $ 9.9^{+0.5}_{- 0.5}$ & $-3.01$ & $ 0.07$ & $-6.30$ & $ 0.08$ & $   61$ & $    3$  \\
 G035.57$-$00.03 & $  91.4_{-  4.9}^{+  4.4}$ & $  -6.2$ & $   7.6$ & $  -5.7$ & $  11.8$ & $   3.3$ & $   6.8$  & $10.2^{+0.6}_{- 0.6}$ & $-3.02$ & $ 0.13$ & $-6.08$ & $ 0.18$ & $   53$ & $    3$  \\
 G032.74$-$00.07 & $  97.6_{-  6.6}^{+  5.5}$ & $   3.9$ & $  12.9$ & $ -19.4$ & $  20.2$ & $   8.7$ & $  13.9$  & $10.6^{+0.8}_{- 0.8}$ & $-3.15$ & $ 0.27$ & $-6.10$ & $ 0.29$ & $   37$ & $   10$  \\
 G020.08$-$00.13 & $ 130.9_{-  3.2}^{+  2.6}$ & $  -7.1$ & $   5.7$ & $   0.3$ & $  17.9$ & $  -5.3$ & $   8.7$  & $12.7^{+0.6}_{- 0.6}$ & $-3.14$ & $ 0.14$ & $-6.44$ & $ 0.16$ & $   41$ & $    3$  \\
 G019.60$-$00.23 & $ 131.3_{-  3.9}^{+  3.1}$ & $  -9.8$ & $   6.0$ & $  -8.2$ & $  19.8$ & $  -4.8$ & $   9.7$  & $12.6^{+0.7}_{- 0.7}$ & $-3.11$ & $ 0.16$ & $-6.36$ & $ 0.17$ & $   41$ & $    3$  \\
\hline
\enddata
\tablecomments{{Peculiar motion for sources listed in Table \ref{table:parallax}.} Column 1 lists the source names nominated with Galactic coordinates. {The sources are sorted in increasing Galactic azimuth as listed in Column 2.} Columns 3, 4, and 5 list the peculiar motions toward the Galactic center, in the direction of Galactic rotation, and toward the north Galactic pole, respectively. Columns 6--9 list the distances, proper motions in the eastward and northward directions, and LSR velocities, respectively. The Galactic ``Univ'' model and solar motions
found by \citetalias{2019ApJ...885..131R} were used to calculate the peculiar motions.  }
\end{deluxetable*}

\clearpage
\bibliography{shuaibo_bibtex}{}
\bibliographystyle{aasjournal}

\end{document}